\documentclass[epj]{webofc}
\usepackage[varg]{txfonts}   
\def\un#1{\,{\rm #1}}

\woctitle{XLV International Symposium on Multiparticle Dynamics}
\begin{document}
\title{Evidence for Non-Exponential Differential Cross-Section of  \\
pp Elastic Scattering at Low $|t|$ and $\sqrt{s} = 8$ TeV by TOTEM 
}
%
%

\author{T. Cs\"org\H{o} \inst{1,2}\fnsep\thanks{\email{tcsorgo@cern.ch}} for the TOTEM Collaboration}

\institute{
Wigner Research Centre for Physics, Budapest, Hungary
\and
KRF, Gy\"ongy\"os, Hungary
          }

\abstract{%

Recently published and preliminary results of the TOTEM experiment are
presented, emphasizing a recent discovery of a non-exponential behaviour of
the differential cross-section of elastic proton-proton scattering, that
TOTEM measured with an unprecedented precision at the centre-of-mass energy
$\sqrt{s} = 8$ TeV based on a high-statistics data sample obtained with the
$\beta_* = 90$ m optics of CERN LHC. Both the statistical and systematic
uncertainties remained below 1\% , except for the t-independent contribution
from the overall normalisation.  This measurement allowed TOTEM  to exclude
a purely exponential differential cross-section in the range of
four-momentum transfer squared $ 0.027 < |t| < 0.2$ GeV$^2$ with a
significance greater than 7 $\sigma$.  In this context we also highlight the
innovative TOTEM recalibration of LHC optics, that used elastic scattering
data measured by the world's largest and most complex Roman Pot detector
system, and discuss recent preliminary TOTEM data on the Coulomb-Nuclear
interference region with its physics implications.

}
\maketitle
\section{Introduction}
\label{sec-1}


The TOTEM experiment at CERN LHC specializes in forward physics
measurements, that deal with colorless exchange, including elastic
proton-proton (pp) scattering, single and double diffractive pp scattering
~\cite{Anelli:2008zza,Antchev:2013hya}, and has a program to measure central
exclusive production in collaboration with the CMS experiment, the CMS-TOTEM
Precision Proton Spectrometer (CT-PPS) project ~\cite{Albrow:2014ctpps}.

The main focus of this conference contribution is recent TOTEM the discovery
of a non-exponential behaviour of the differential cross-section of elastic
pp scattering ~\cite{Antchev:2015zza} at low $|t|$ and at $\sqrt{s} =  8$
TeV energy of CERN LHC, which is put into context of other recent TOTEM
results, in particular the recalibration of LHC optics from TOTEM data
~\cite{Antchev:2014voa} and preliminary results on the observation of
Coulomb-nuclear interference at very low four-momentum transfer at LHC
~\cite{Mario:2015LHCC}.

\section{Earlier results on the low-$|t|$ behaviour of $d\sigma/dt$ of elastic pp scattering}
\label{sec-2}

Let us first summarize what has been known before 2015 about this low-|t|
region of elastic scattering in $pp$ and $p\overline{p}$ scattering.
Traditionally, the differential cross-section of elastic scattering is
characterized in the low-|t| region, before the diffractive minimum, by an
approximately exponential behaviour, $\exp{(-B |t|)}$, where $B\equiv
B(s,t)$ is called the slope parameter, where $s = (p_1 +p_2)^2 $ and  $ t =
(p_1-p_3)^2 $ are Mandelstam variables called squared center of mass energy
and the squared four-momentum transfer. 


Experimental data reported before 2015 on $pp$ and $p\overline{p}$ elastic
scattering were consistent with a simple exponential behaviour, $B(s,t) =
B(s)$ independent of $t$ in the low $|t|$ region, before the vicinity of
$t_{dip}$, the region of the diffractive minimum.  The slope parameter
$B=B(s) $ was measured to increase with increasing $\sqrt{s}$, corresponding
to the shrinking of the diffractive cone and the increase of the total
elastic cross-section of proton-proton scattering with increasing center of
mass energies.
Nevertheless, some of the earlier experiments reported hints of deviations
from such a simple exponential behaviour. For example, small angle $pp$
elastic scattering at 460 $< \sqrt{s} <$ 2900 GeV energies indicated a
marked change of the slope parameter as a function of $t$ in the $ |t |= 0.1
$ GeV$^2$ region as early as in 1972 ~\cite{Barbiellini:1972ua}. A decade
later, an indication of a break in the slope parameter $B$ was observed in
$p\overline{p}$ reactions, as a function of $t$, at the ISR energy of 52.8
GeV in ref.  ~\cite{Ambrosio:1982zj}.  Other measurements at ISR found a
simple exponential behaviour, $B\ne B(t)$ at 53 and 62 GeV ISR energies,
however, ref.~\cite{Breakstone:1984te}
found an indication of a change of the slope parameter at $|t|$ =
0.15 GeV$^2$ at $\sqrt{s} = 31$ GeV colliding energies, and found
that the data are better described by a Gaussian $\exp(-B |t| - C t^2)$ than
a simple exponential  $\exp(- B|t|)$ shape.  These early results
were limited by statistics and systematic effects  and the significance of
these indications of non-exponential behaviour did not reach the discovery,
$5\sigma$ level. In measurements of elastic $p\overline{p}$ scattering at
Tevatron energies, no indications of a possible non-exponential behaviour
were found.
The first results on the differential cross-section of elastic $pp$
scattering at LHC energies were reported by TOTEM in the dip
region~\cite{Antchev:2011zz}, but the TOTEM measurement was soon extended to
the low-$|t|$ region using a special LHC run~\cite{Antchev:2013gaa}. Both
measurements were consistent with an exponential behaviour in the
diffraction cone, and the measured slope parameter in the
low-$|t|$ region was consistent with the $B$ value determined from the data
closer to the dip region. However, a high statistics low-$|t|$ data-set from a special LHC run allowed TOTEM to establish the non-exponential behaviour
of the differential cross-section of elastic $pp$ scattering with a greater than 7 $\sigma$ significance at $\sqrt{s} = 8$ TeV LHC colliding energies
~\cite{Antchev:2015zza}.


The experimental observation of non-exponential low-$|t|$ behavior of the
differential cross-section of elastic $pp$ scattering was actually
anticipated by a number of theoretical models.  
To explain the first experimental indications of a non-exponential behaviour
at ISR energies, ref.~ \cite{CohenTannoudji:1972gd} suggested that the
corresponding non-linearity of the vacuum or Pomeron Regge trajectory is due
to a nearby singularity, the two-pion threshold, restricted by t-channel
unitarity requirements. This idea has been applied recently
~\cite{Jenkovszky:2014yea} to explain the non-exponential behaviour of
preliminary TOTEM data at 8 TeV extrapolating fit parameters  from ISR to
LHC energies~\cite{Fagundes:2015vva}.  An independent line of argumentation
was based on multiple diffraction theory of elastic proton-proton scattering
~\cite{Glauber:1984su} where the electromagnetic form-factor of the
colliding protons and a simple phenomenological parameterizaton of the
cluster-averaged parton-parton forward scattering amplitude was shown to
lead naturally to a non-exponential behaviour or a $t$-dependent
slope-parameter $B(t)$.  This general quantum-optical scattering theory is
also able to describe the first TOTEM data if the proton is modelled as a
quark-diquark composite object, $p = (q, d)$ even if the internal structure
of the quark and the diquark is a featureless Gaussian distribution as far
as elastic scattering of protons is concerned. The resulting multiple
diffraction theory formulas for elastic $pp$ scattering indicate a
non-exponential feature both  in the ``springy Pomeron" picture of Grichine~
\cite{Grichine:2014wea} and the Bialas-Bzdak quark-diquark model of $pp$
elastic scattering, ~\cite{Bialas:2006qf}, in particular  when the
non-vanishing nature of the real part of the forward scattering amplitude
and the unitarity constraints are both taken into account
~\cite{Nemes:2015iia}.

\section{Non-exponential behaviour of low-$|t|$ elastic $pp$ scattering}
\label{sec-3}
In this section, the main results of ref.  
~\cite{Antchev:2015zza} 
are summarized and the method of the analysis is outlined.

{\it Detector setup.}
The TOTEM experiment~\cite{Anelli:2008zza,Antchev:2013hya} is located at the LHC interaction point (IP) 5,
together with the CMS experiment. The full experimental apparatus of TOTEM
consists of two inelastic telescopes, T1 and T2, based on Cathode Strip
Chamber and GEM technology, respectively, and a system of Roman Pot
detectors.  Roman Pots are movable beam-pipe insertions that approach the
LHC beam very closely in order to detect particles scattered at very small
angles. Currently, 26 Roman Pot detectors belong to TOTEM and the CT-PPS
project: this is the largest and most complex Roman Pot  detector system that
has ever been operated at a collider.
In the measurement of the non-exponential shape of the differential cross-section of elastic $pp$ scattering, only the Roman Pot (RP) system, the
sub-detector relevant for elastic scattering measurement, was utilized.

The RP-s are organised in two stations placed symmetrically around the IP:
one on the left side (in LHC sector 45), one on the right (sector 56). Each
station is formed by two units: near ($214$ m from the IP) and far
($220$ m). Each unit includes three RPs: one approaching the beam from
the top, one from the bottom and one horizontally. Each RP hosts 10
``edgeless" silicon strip sensors with a pitch of $66\un{\mu m}$, each
having a strongly reduced insensitive edge of a few tens of micrometers
facing the beam.  The sensors are equipped with trigger-capable electronics.
Since elastic scattering events consist of two anti-parallel protons, the
detected events can have two topologies, called diagonals: 45 bottom -- 56
top and 45 top -- 56 bottom.

{\it Data-set}.  The measurement presented here is based on data taken in
July 2012, during the LHC fill number 2836 at the centre-of-mass energy $\sqrt s = 8\un{TeV}$. The vertical RPs were
inserted at a distance of $9.5$ times the transverse beam size, $\sigma_{\rm
beam}$. Initially two, later three colliding bunch-pairs were used, each
with a typical population of $8\cdot10^{10}$ protons, corresponding to an
instantaneous luminosity of about $10^{28}\un{cm^{-2}s^{-1}}$ per bunch. The
main trigger required a coincidence between the RPs in both arms, combining
the near and far units of a station in \textit{OR} to ensure maximal
efficiency. During the about $11\un{h}$ long data-taking, a luminosity of
$735\un{\mu b^{-1}}$ was accumulated, giving $7.2\cdot 10^6$ tagged elastic
events.

\begin{figure}
\centering
\includegraphics[width=2 truein,clip]{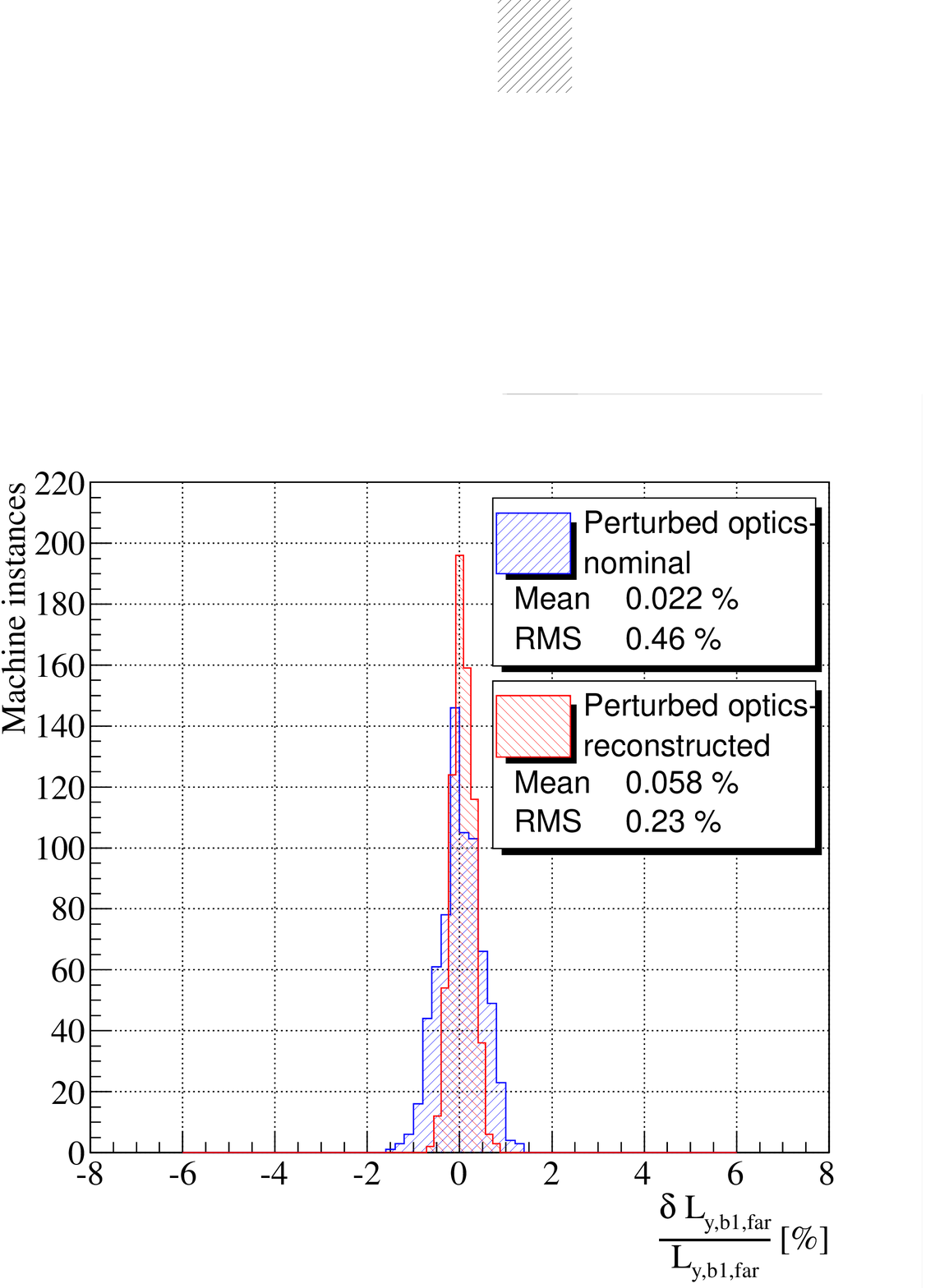}
\includegraphics[width=2 truein,clip]{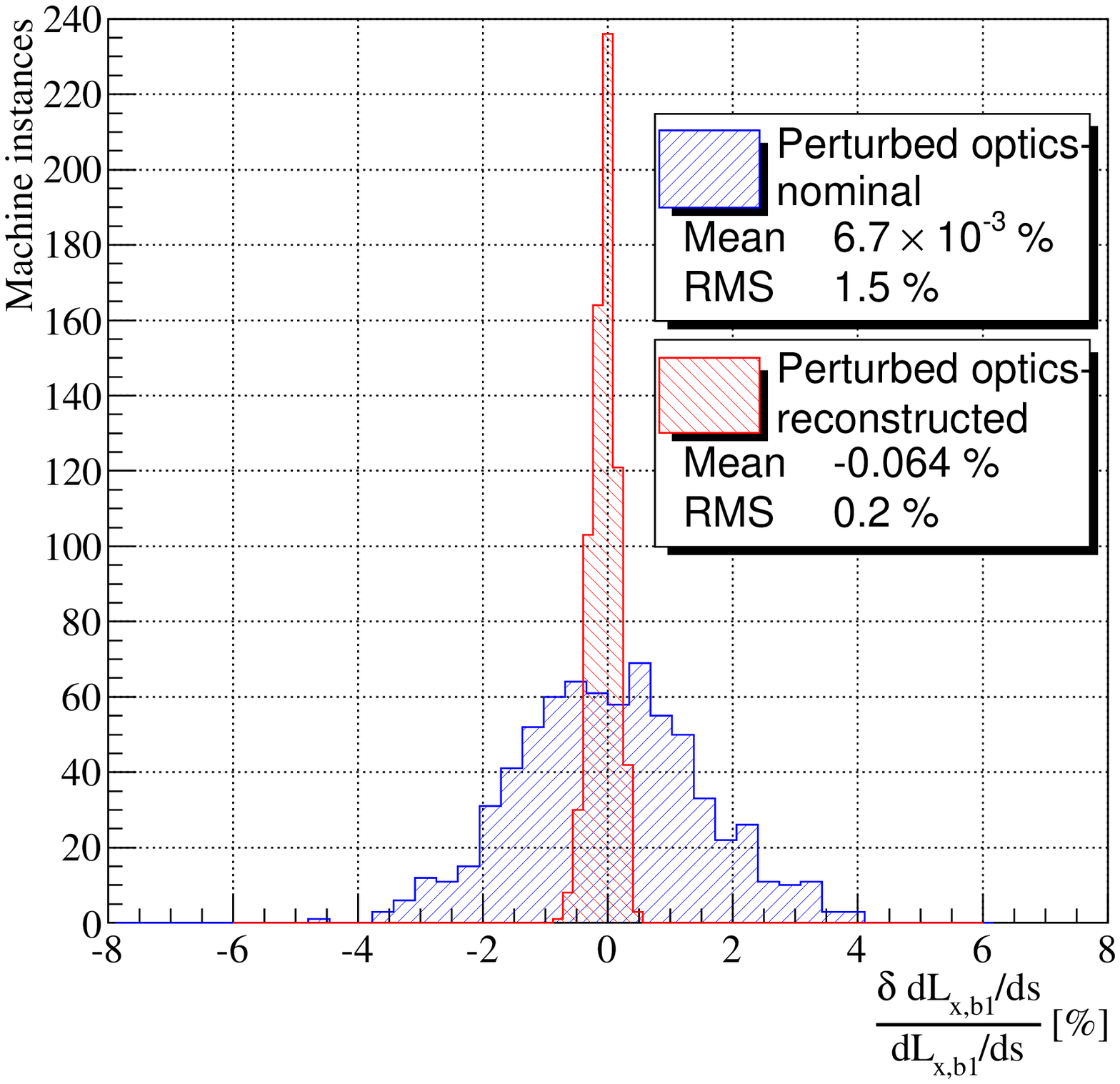}
\caption{Monte-Carlo error distribution of $\beta_* = 90$ m LHC Optical Functions $L_y$ and $dL_x/ds$ for E = 4 TeV before and after the estimation of the LHC Optics from TOTEM $pp$ elastic scattering data~\cite{Antchev:2014voa}.
 }
\label{fig-1}       
\end{figure}

{\it LHC Optics Determination at $\beta_* = 90 $ m}.  For the determination
of the total cross-section and the differential cross-section of elastic
$pp$ collisions with unprecedented precision by TOTEM, a precise
experimental  control of the LHC optics was necessary, which was achieved
from the recalibration of the LHC optics at IP5 using measurement of elastic
$pp$ scattering with RP-s.  For the measurement of the non-exponential shape
of $d\sigma/dt$, a special LHC optics with $\beta^* = 90\un{m}$ was used,
with essentially the same characteristics as at $\sqrt s = 7\un{TeV}$.  In
the vertical plane, this optics features parallel-to-point focusing ($v_y
\approx 0$) and large effective length $L_y$. In the horizontal plane, the
almost vanishing effective length $L_x$ simplifies the separation of elastic
and diffractive events: any sizeable horizontal displacement must be due to
a momentum loss $\xi$. Figures ~\ref{fig-1}       indicate, that TOTEM
achieved a precise control of LHC imperfections with perturbed LHC optics
and recalibration from data at IP5: Monte-Carlo simulations indicate that
the nominal uncertainties of LHC optics have  been reduced by factors of 2 -
10, see Refs.~\cite{Antchev:2014voa}, 
~\cite{Antchev:2015zza} 
for details of this innovative TOTEM method.

{\it Analysis steps}.
The precise control of LHC optics, together with kinematic cuts that
are based on the physics of elastic scattering
were needed to measure precisely the differential cross-section.
The applied analysis steps included detector alignment, 
optics recalibration, elastic event selection based on kinematic cuts,
resolution unfolding, acceptance correction, background subtraction,
detection  and efficiency correction, corrections for angular resolution,
normalization and binning.  

Compatible results have been obtained from the data originating from
different bunches, different diagonals and different time periods. In
addition, the complete analysis chain has been applied in two independent
analysis implementations, yielding compatible results.
The details of these steps of the  analysis are described 
in ref. ~\cite{Antchev:2015zza}.

\begin{figure}
\centering
\includegraphics[width=4.5 truein,clip]{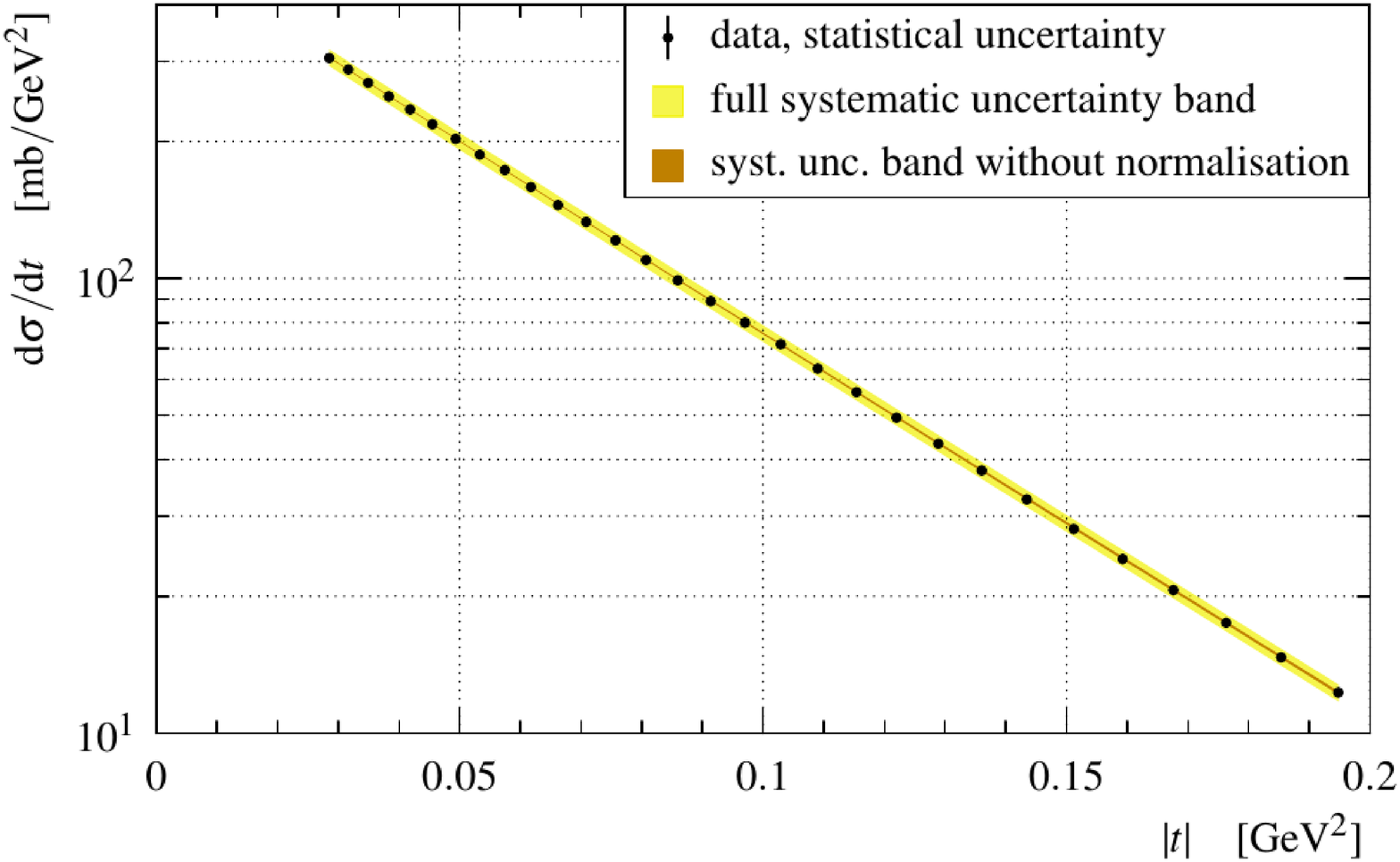}
\includegraphics[width=4.5 truein,clip]{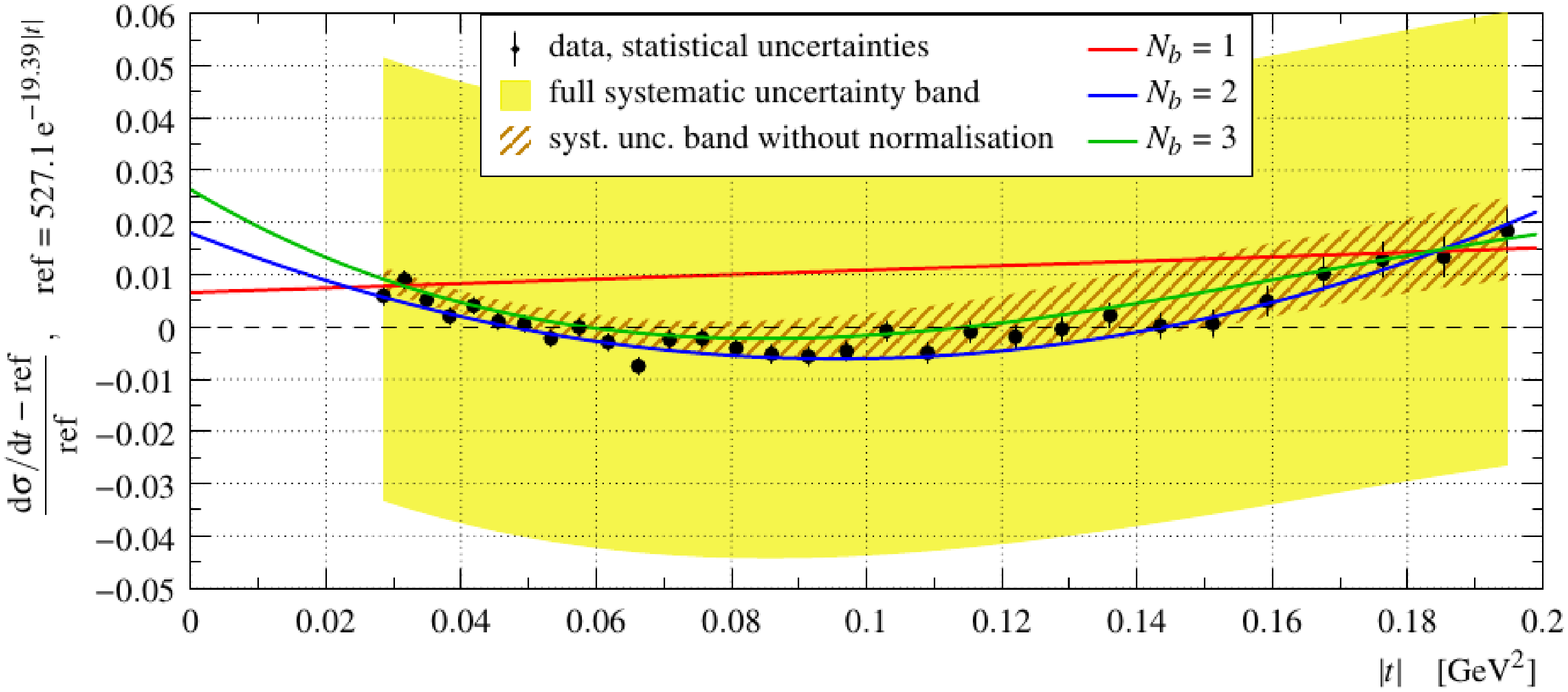}
\caption{
The {\it top panel} shows the differential cross-section of elastic $pp$
scattering at 8 TeV as measured by TOTEM at CERN LHC.  The black dots
represent TOTEM data with statistical uncertainty bars. The widest error
band (yellow) corresponds to full systematic uncertainty, including the
normalization error. The narrow band around the data points indicate all
systematic errors except the overall normalization uncertainty. Both bands
are centered around the fit curve with $N_b = 3$.  The {\it bottom panel}
shows the same data points, however, now the relative deviation from an
exponential reference distribution is shown (see the vertical axis).  
From ref.~\cite{Antchev:2015zza}.
}
\label{fig-2}       
\end{figure}

{\it Results on the non-exponential behaviour}.
The differential cross-section of elastic $pp$ scattering at 8 TeV
is indicated on the top panel of Figure ~\ref{fig-2},
tabulated in Table 3 of ref. ~\cite{Antchev:2015zza}.
The binning of this plot is optimized so that the
bin size corresponds to 1$\sigma$ of the resolution in $t$.  
In order to visualise small deviations from the leading pure-exponential
behaviour, the bottom panel of Figure~\ref{fig-2} shows also
the relative difference of the
cross-section from a reference exponential (pure exponential fit using
statistical uncertainties only). This plot indicates the
non-exponentiality of the data: pure exponentials would look like  (nearly)
linear functions in this kind of representation.

To study the detailed behaviour of the differential cross-section, a series of fits has been made using the parametrisation:
\begin{equation}
\label{eq:fit param}
\frac{\mathrm{d}\sigma}{\mathrm{d} t}(t) = 
\frac{\mathrm{d} \sigma}{\mathrm{d} t}
(t=0) \, \times \, 
\exp\left( \sum\limits_{i = 1}^{N_{b}} b_i\, t^i \right) \ ,
\end{equation}
which includes the exponential fit as a special case for ($N_b = 1$) and its
straight-forward extensions ($N_b = 2, 3$).  The fits have been performed by
the standard least-squares method, using the covariance matrix including
both the statistical and systematic components.  For a detailed description
of the determination of the systematic errors, as well as for a discussion
on the significance of these observations, see ref.~\cite{Antchev:2015zza}.
Two independent lines of analysis indicate that the significance of the
non-exponential behaviour in these TOTEM measurements is greater than
7$\sigma$. Using the refined, non-exponential parameterizations to
extrapolate to the optical point, $t= 0$, yields total cross-section valued
compatible with earlier TOTEM determinations, in all cases neglecting the
effects due to the Coulomb interactions.

\begin{figure}
\centering
\includegraphics[width=3 truein,clip]{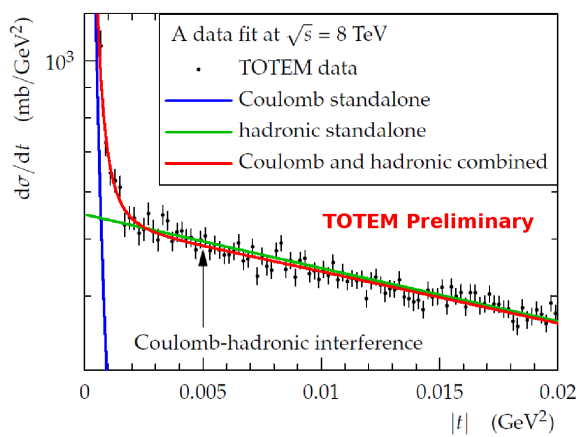}
\caption{
TOTEM preliminary results with a special $\beta_* = 1000 $ m run, Roman Pots
approaching the beam as close as 3.5 $\sigma$ probe the Coulomb-Nuclear
Interference (CNI) region. 
From~\cite{Mario:2015LHCC}.
 }
\label{fig-3}       
\end{figure}

{\it Preliminary TOTEM Results on the Coulomb-Nuclear Interference.}
\label{sec-4}
In a preliminary analysis, TOTEM investigated if the origin of the
non-exponentiality at low-$|t|$,  still dominated by the hadronic
interactions, can be due to Coulomb interference effects or not. This was
achieved with a measurement of the elastic scattering in the Coulomb-Nuclear
Interference (CNI) region.
The preliminary TOTEM data, shown in Fig.~\ref{fig-3},
extend the four momentum transfer range of the TOTEM measurements 
down to $|t|\approx 6 x 10^{-4}$ GeV$^2$. These  results 
indicate that the non-exponential behaviour of $d\sigma/dt$ around 0.1
GeV$^2$ is not due to Coulomb interference effects, but is a property of the
hadronic amplitude ~\cite{Mario:2015LHCC}, 
which allows TOTEM to exclude the simplified West-Yennie interference
formula ~\cite{West:1968du}. 
The constraining power of this TOTEM preliminary
analysis on the hadronic phase allows us to study the impact parameter
dependence of elastic $pp$ scattering and to measure the parameter $\rho$. 
The results of this TOTEM preliminary analysis confirm 
earlier total $pp$ scattering cross-sections measurements of TOTEM,
that neglected the Coulomb-Nuclear Interference effects.

{\it Acknowledgments:}
T. Cs. would like to thank the Organizers of this Symposium for their
support and for an inspiring and useful meeting.  This work was supported by
the Magnus Ehrnrooth foundation (Finland), the Waldemar von Frenckell
foundation (Finland), the Academy of Finland, the Finnish Academy of Science
and Letters (the Vilho, Yrj\"o and Kalle V\"ais\"al\"a Fund), 
and the OTKA grant NK 101438 (Hungary). 


\vfill\eject

\end{document}